\documentclass[manuscript]{acmart}

\AtBeginDocument{%
  \providecommand\BibTeX{{%
    \normalfont B\kern-0.5em{\scshape i\kern-0.25em b}\kern-0.8em\TeX}}}

\begin{document}

\title{Defining a canonical unit for accounting purposes}

%% CHECK THE ORDER
\author{Fabio Andrijauskas}
\email{fandrijauskas@ucsd.edu}
\orcid{0000-0002-1254-8570}
\affiliation{%
  \institution{University of California - San Diego}
  \streetaddress{9500 Gilman Dr}
  \city{La Jolla}
  \state{CA}
  \country{US}
  \postcode{92093}
}
\author{Igor Sfiligoi}
\email{isfiligoi@sdsc.edu}
\orcid{0000-0002-9308-5327}
\affiliation{%
  \institution{University of California - San Diego}
  \streetaddress{9500 Gilman Dr}
  \city{La Jolla}
  \state{CA}
  \country{US}
  \postcode{92093}
}

\author{Frank W\"urthwein}
\email{fkw@ucsd.edu}
\orcid{0000-0002-8115-7573}

\affiliation{%
  \institution{University of California - San Diego}
  \streetaddress{9500 Gilman Dr}
  \city{La Jolla}
  \state{CA}
  \country{US}
  \postcode{92093}
}

\begin{abstract}
Compute resource providers often put in place batch compute systems to maximize the utilization of such resources.
However, compute nodes in such clusters, both physical and logical, contain several complementary resources, with notable examples being CPUs, GPUs, memory and ephemeral storage. User jobs will typically require more than one such resource, resulting in co-scheduling trade-offs of partial nodes, especially in multi-user environments. When accounting for either user billing or scheduling overhead, it is thus important to consider all such resources together.
We thus define the concept of a threshold-based "canonical unit" that combines several resource types into a single discrete unit and use it to characterize scheduling overhead and make resource billing more fair for both resource providers and users.
Note that the exact definition of a canonical unit is not prescribed and may change between resource providers. Nevertheless, we provide a template and two example definitions that we consider appropriate in the context of the Open Science Grid.
\end{abstract}


\maketitle

\section{Introduction}

Many research computing providers have to contend with user demand that exceeds the available resources under their control.
Batch compute systems are thus the preferred resource management system, as they allow for both high resource utilization and fair sharing in multi-user scenarios. 

High utilization and low user wait times are, however, often at odds with each other, especially when user jobs require only a fraction of the resources available in any compute node, either physical or logical.
Ideal packing of jobs may result in some classes of jobs never being scheduled. Most resource schedulers thus balance job priorities with resource utilization. It is in everyone's interest to keep resource utilization high, as that allows for most compute jobs to complete over an extended period of time. There are many ways to achieve that, both on the user and provider side, so proper accounting of resource utilization and scheduling overhead is essential for guiding the optimization process.

Accounting systems have been part of the batch ecosystem for a long time, but they usually measure each resource type as an independent metric. Different resource types, e.g., CPUs, GPUs, memory and ephemeral storage, aka scratch space, are however not independent. Every user job needs several of them, at the very least some CPU cores and some amount of memory, and cannot be scheduled if any of them is not available. While independent-resource accounting systems provide reasonable means to measure the usage of such resources, they fall significantly short in characterizing the scheduling overhead, i.e., why resources are not used at any point in time.

We thus define the concept of a \textbf{canonical unit}, which we use to define the \textbf{true overhead}, i.e., the per-node difference between available and allocatable resources. The \textit{canonical units} are expressed as integer numbers and are computed using a threshold-based combination of all the resources of interest to the cluster operator. We then proceed to describe how these concepts help understand both scheduling trade-offs and improve the billing policies.

Note that the exact definition of a \textit{canonical unit} is not prescribed and may change between resource providers. Nevertheless, in this paper we provide a template definition, alongside two example definitions that we consider appropriate in the context of the Open Science Grid (OSG) \cite{Pordes_2007}, one for CPU-only nodes and one for GPU nodes.

Finally, we want to clarify that own main focus was in the context of improving the accounting capabilities of OSG's GRACC \cite{Retzke_2017} in the context of glideinWMS \cite{5171374}. Nevertheless, the results presented are general and should be applicable to many other systems and deployments.

\section{The partial node scheduling problem}

In recent years compute hardware has become faster mostly by means of adding parallelism, with relatively minor improvements in single threaded performance. While it is still possible to buy and operate CPUs with only a few cores, many-core CPUs and massively parallel accelerators, like GPUs, provide a more cost effective solution.

Unfortunately, parallel software development is non trivial \cite{4484943}, so a large fraction of scientific applications cannot make effective use of all the resources in such beefy nodes. In order to properly support such applications, the batch scheduling system has to partition the node among several independent user jobs.

There are several possible ways to partition nodes, each with its own benefits and drawbacks:
\begin{enumerate}
    \item  At one extreme, one can procure identical hardware, evenly partition each physical node in many smaller logical nodes and then schedule such logical nodes as indivisible units. The major drawback of such system is the inability of properly serving large memory and parallel-ready applications, as there are no large logical nodes in the system.
    \item At the other extreme, one can allow for nodes of different sizes and partition each node during the matchmaking stage. This maximizes the flexibility of the system during the bootstrap phase, but can fast result in sufficient fragmentation that only the smallest jobs ever get matched. One thus has to employ at the node-level advanced scheduling techniques, like reservations, typical of many-node HPC deployments \cite{osti_15002962}.
    \item The glideinWMS system employs an in-between solution. In simplified terms, it periodically partitions the nodes using one of the few pre-set configurations with a well defined lifetime, based on the current needs of the jobs in the queues. Those partitions are then further split as-needed during the matchmaking phase, with the remaining logical node lifetime being the only hard limit. As a side effect of this policy, the system performs a forced defragmentation at each higher-level partitioning interval.
\end{enumerate}

In a typical multi-user environment, where queued jobs span a wide range of requirements in multiple dimensions, all partitioning schemes will result in some unused resources at least part of the time. In the first extreme, it will be due to over-provisioning of resources compared to job needs, due to the rigid nature of the resource splitting. In all other cases, the resources will be occasionally idled due to scheduling constraints. Given the many scheduling options, it is desirable to properly understand this resource under-utilization, a.k.a the \textit{scheduling overhead}.

Note that we are strictly restricting ourselves to the difference between the sum of job requirements and available resources. We do not consider the actual use of those resources by the processes inside the user jobs themselves. While we acknowledge that's important, too, it is outside the scope of this document.

The currently typical accounting practice is to keep track of each resource allocation separately and to average the values over many nodes. While this does provide a great metric for how effectively those resources are being used, it does not provide enough information about the causes of the scheduling overhead.

\begin{figure}[H]
         \centering
         \includegraphics{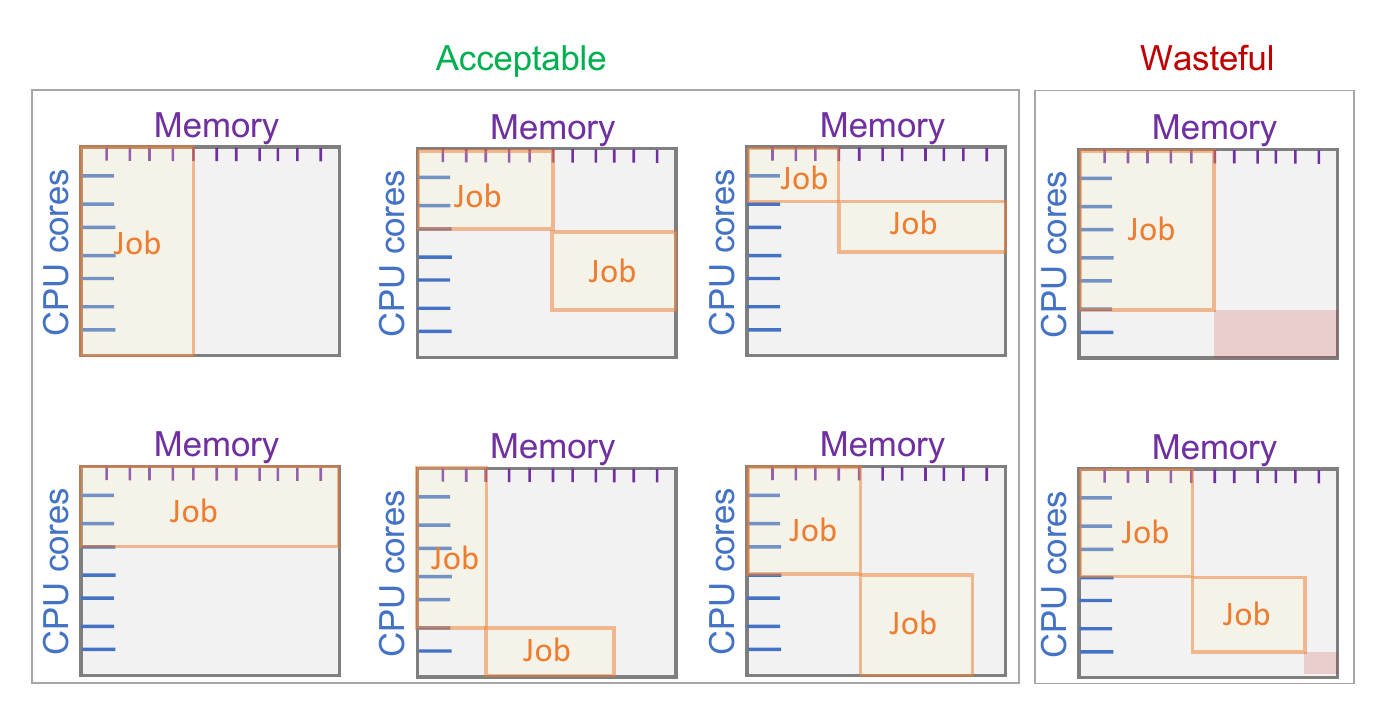}
         \caption{Examples of partial node occupancy.}
         \label{nodes1}
\end{figure}

The main problem of existing accounting systems is their inability to discriminate between situations where only a subset of resources on a node is underutilized and situations where all of the resources on a node are under-utilized. Since jobs need all of their requirements to be satisfied in order to run, keeping at least one resource type fully utilized at all times may be the best we can aim for. For example, as shown in Figure \ref{nodes1}, if the users have only a mix of big-memory few-thread jobs and large-thread small-memory jobs, it may be perfectly legit that some nodes have all memory in use but only a few cores busy, while the others have all the cores in use but a large fraction of the memory is not in use. What we want to avoid is having both some cores and some memory unallocated at the same time. Unfortunately, an accounting system measuring CPU core and memory allocations independently at the cluster level does not provide such insight.

\section{Defining the Canonical Unit}

In order to address the problem outlined in the previous section, we thus define the concept of a "\textit{canonical unit}", which we will use as the largest discreet count representing "\textit{true overhead}", i.e., the per-node difference between available and allocated resources. The \textit{canonical units} are expressed as integer numbers and are computed using a threshold-based combination of all the resources of interest to the cluster operator. 

Note that we do not provide a prescribed way to define the thresholds, leaving that to the discretion of the cluster operator, and even allowing for different thresholds to be used for different partitions of the cluster. While we acknowledge that this will make comparison between different clusters almost impossible, we believe such flexibility is needed to maximize the insight a cluster operator gets and to make the concept future-proof. Furthermore, nothing prevents a cluster operator to account using different \textit{canonical unit} thresholds, e.g., one targeted and one more standardized, so in the following sections we propose a few example threshold definitions that we will consider for future more formal standardization.

\begin{figure}[H]
         \centering
         \includegraphics{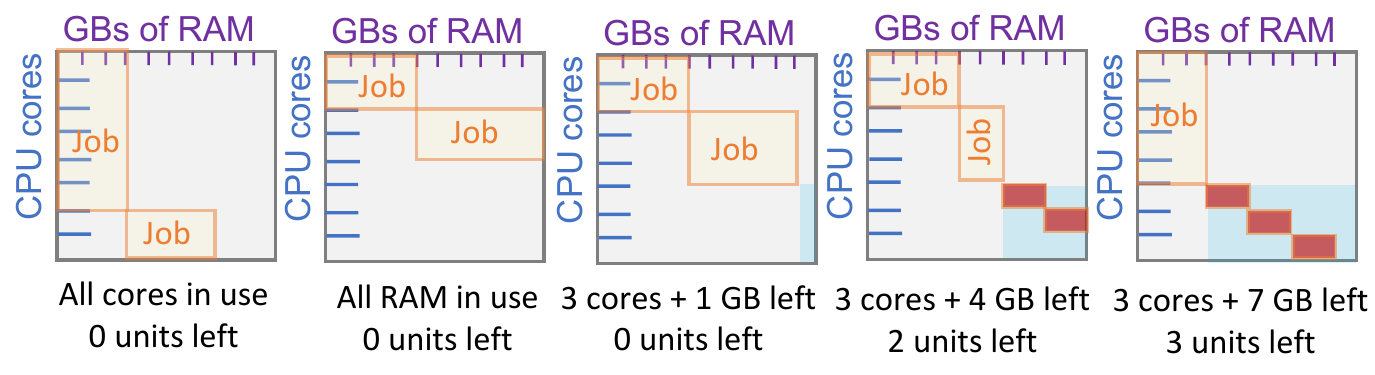}
         \caption{Example use cases of true overhead using canonical units, 
                   using "1 CPU core and 2 GB of memory" as the canonical unit thresholds. }
         \label{cu2}
\end{figure}

Nevertheless, in order to make the text more readable, we use a specific thresholds of "1 CPU core and 2 GB of memory" in Figure \ref{cu2} to present a few examples:
\begin{itemize}
  \item In the case where either all CPU cores or all memory of a node is allocated, we treat that node as having 0 \textit{canonical units} of \textit{true overhead}.
  \item In the case of a node where there are 3 CPU cores and 1 GB of memory that have not been allocated to jobs, we treat that node still as having 0 \textit{canonical units} of \textit{true overhead}. I.e., we ignore any leftovers that do not meet the thresholds.
  \item In the case of a node where there are 3 CPU cores and 4 GB of memory that have not been allocated to jobs, we treat that node as having 2 \textit{canonical units} of \textit{true overhead}. I.e., we stop once all the memory is allocated.
  \item In the case of a node where there are 3 CPU cores and 7 GB of memory that have not been allocated to jobs, we treat that node as having 3 \textit{canonical units} of \textit{true overhead}. I.e., we stop when all the CPU cores are in use.
\end{itemize}

\subsection{Interpreting true overhead accounting}

Using the \textit{canonical unit}-based \textit{true overhead} for accounting, it is easy to see that only the last two example scenarios are undesirable. The same cannot be said for traditional, independent resource accounting, which would have flagged all of them as undesirable.

The difference is even more pronounced when looking at aggregate values across many nodes. In the context of Figure \ref{cu2}, the traditional, independent resource accounting systems would be unable to distinguish between a mix of the two leftmost use cases and a mix of the two rightmost use cases, unlike a \textit{canonical unit}-based \textit{true overhead} accounting system.

To maximize the interpretative power of \textit{canonical unit}-based \textit{true overhead} accounting systems we also recommend the use of histograms for aggregation purposes. On top of a pure quantitative metric, a \textit{true overhead} histogram would also allow for cross correlation with job queues, enabling detection of situations where resources are under-utilized while there are eligible jobs waiting to be run. The same would be much harder to achieve using the independent resource accounting, due to the interpretabillity limits of multi-dimensional histograms.

As an example, the two rightmost use cases in Figure \ref{cu2} should ideally only happen when there are no jobs waiting in the queues that request less than 2 CPUs and 4 GB of memory, i.e. 2 \textit{canonical units}. If instead we can find such jobs, that would indicate scheduling overhead that is likely tied to advanced scheduling throttles, like reservations and defragmentation.

Note that we intentionally avoid the topic of job queue monitoring and accounting implementation. While they likely would benefit from incorporating the notion of \textit{canonical unit}, we consider that topic outside the scope of this paper, although we may address it in future research.

\subsection{Usage accounting and true overhead accounting interplay}

The proposed \textit{canonical unit} is essential for proper accounting of \textit{true overhead}, but cannot be used for resource usage accounting. As mentioned in the previous section and in the examples, most systems serve jobs that have a wide range of requirements, some large in one dimension, others in a completely different one, making them a bad fit for the rigid \textit{canonical unit} framework. We are thus not proposing any changes to the resource usage accounting systems.

Moreover, we envision resource usage accounting systems to coexist with the \textit{true overhead} accounting systems, as they provide complementary benefits. While a \textit{true overhead} accounting system provides insight into scheduling overhead proper, the main goal of any batch system operator would still be to maximize the resource usage for all the available resources.

\subsection{Using true overhead for billing purposes}

Most batch system operators charge their users for the resources they consume, using a combination of allocations, priorities, quotas and/or monetary compensation. When nodes cannot fully be utilized due to a mismatch between hardware procurement and job request strategies, someone has to take a loss.

Using a pure resource usage accounting system users are only charged for the resources that they requested, so the loss is fully shouldered by the resource providers, since any leftovers logically just go to waste. While users arguably like this situation, it may not be considered fair by resource providers, as they may be foregoing ideal packing to improve user experience in the form of lower latencies and proper fair-share scheduling.

An alternative billing strategy is to always charge the cost of the whole node to the users whose jobs run there, no matter how many resources were requested by those jobs. While this strategy would be ideal for resource provides, it is arguably not fair to the users, as scheduling policies and related overheads are outside their control sphere.

We thus define an intermediate billing scheme that we believe is fair to both users and resource providers. In this alternate billing scheme, the \textit{true overhead} of each node is never charged to the users, i.e. it is realized as a loss by the resource providers. Whatever remains is then proportionally charged to the users whose jobs run on that node.

As en example, let's consider the jobs in Figure \ref{cu2}:
\begin{itemize}
  \item In the leftmost use case, there is no \textit{true overhead} as all CPUs are in use. But 30\% of the memory was not requested. So the two jobs will be billed for CPUs at 100\% rate and for memory at 143\% rate (\(\frac{10}{7}\)).
  \item Similarly, the next use case has no \textit{true overhead}, but 50\% of the CPU cores are unused. The two jobs will thus be billed for CPUs at 200\% rate and for memory at 100\% rate.
  \item The rightmost use case has 3 \textit{canonical units} of \textit{true overhead}, so those 3 CPU cores and 6 GBs of memory will not be considered for billing purposes. That still leaves 10\% of unused memory, so the job will be billed for CPUs at 100\% rate and for memory at 111\% rate (\(\frac{10}{9}\)).
\end{itemize}

We believe the above billing scheme is fair for the resource providers, as they can control \textit{true overhead} through provisioning and scheduling policies. 

And even though users may get charged more than what they requested, we believe the billing is still fair to the users, assuming that the billing policy was properly disclosed. The billed resources do not include a scheduling overhead and have a tangible cost to the resource providers, so we believe that it is fair to ask users to pay for them. At fixed \textit{true overhead}, resource providers do not benefit from different packing mechanisms, so we argue that it is fair to proportionally spread the cost among all running jobs.

\section{Two canonical unit definitions applicable to the Open Science Grid}

Based on our experience, resource providers typically procure hardware using a balanced approach between different resource types. The \textit{canonical unit} thresholds should thus represent such procuring logic.

In the Open Science Grid (OSG) communities, the two currently most popular procurement strategies seem to be:
\begin{itemize}
    \item For CPU-only nodes, procure 2 GB of memory for each CPU core.
    \item For GPU nodes, procure as many GPUs that fit in the node, with everything else being almost an afterthought.
\end{itemize}

This thus translates in two distinct definition of \textit{canonical unit} thresholds for the two partitions:
\begin{enumerate}
    \item For CPU-only nodes, we suggest using "1 CPU core and 2 GB of memory" as \textit{canonical unit} thresholds.
    \item For GPU nodes, we suggest using "1 GPU chip" as the only \textit{canonical unit} threshold.
\end{enumerate}

\section{Summary and conclusion}

Understanding scheduling overhead in partial node scheduling setups is highly desirable, as it can help improve resource utilization without excessively affecting user experience. Traditional accounting systems do not provide the necessary information, so we propose a complementary accounting system based on the threshold-based multi-resource \textbf{canonical unit}. 

The use of the \textit{canonical unit} can also improve the fairness of resource billing schemes, properly distributing the cost of under-utilization between users and resource providers.

\begin{acks}

This work was partially supported by the National Science Foundation (NSF) grant OAC-2030508.
\end{acks}

\bibliographystyle{ACM-Reference-Format}
\bibliography{bibliography}

\end{document}